\documentclass{article}
\usepackage{dcase2022_techrep,amsmath,graphicx,url,times,booktabs, tabularx}
\usepackage[T1]{fontenc}
\usepackage{caption}
\usepackage{amsmath}
\usepackage{amssymb}

\title{EFFICIENT AUDIO CAPTIONING TRANSFORMER WITH PATCHOUT AND TEXT GUIDANCE}

\name{Thodoris Kouzelis$^{1,2}$,
      Grigoris Bastas $^{1,2},$
      Athanasios Katsamanis$^{2}$, 
      Alexandros Potamianos$^{1}$,
      }

\address{$^1$ Institute for Language and Speech Processing, Athena Research Center, Athens, Greece,\\
   \{theodoros.kouzelis, g.bastas, nkatsam\}@athenarc.gr\\          
        $^2$ School of ECE, National Technical University of Athens, Athens, Greece, \\
        \{potam\}@central.ntua.gr
 }

\begin{document}

\ninept
\maketitle

\begin{sloppy}

\begin{abstract}
Automated audio captioning is multi-modal translation task that aim 
to generate textual descriptions for a given audio clip.
In this paper we propose a full Transformer architecture that utilizes \textit{Patchout} as proposed in \cite{passt},  significantly reducing the  computational complexity and avoiding overfitting.
The caption generation is partly conditioned on textual AudioSet tags extracted by a pre-trained classification model which is fine-tuned to maximize the semantic similarity between AudioSet labels and ground truth captions. To mitigate the data scarcity problem of Automated Audio Captioning we introduce transfer learning from an upstream audio-related task and an enlarged in-domain dataset.
Moreover, we propose a method to apply Mixup augmentation for AAC.
Ablation studies are carried out to investigate how \textit{Patchout} and text guidance contribute to the final performance.
The results show that the proposed techniques improve the performance of our system and while reducing the computational complexity. Our proposed method received the Judges Award at the Task6A of DCASE Challenge 2022.
\end{abstract}


\section{Introduction}
\label{sec:intro}

Automated Audio Captioning (AAC) is a multimodal task that aims to generate textual descriptions for a given audio clip.
In order to generate meaningful descriptions, a method needs to capture the sound events present in an audio clip and identify the spatial-temporal relationships between them.
 It has great practical potential in various applications such as assisting people who are deaf or hard of hearing (DHH).

One of the main challenges of AAC is the lack of sufficient data. To address it, many recent approaches utilize pre-trained models such us PANNs and VGGish networks, improving the final performance \cite{transfer1, transfer2}.
Mei et al. \cite{act} use a pre-trained transformer encoder, as transformers have recently been shown to outperform CNNs in audio classification tasks \cite{ast}. One limitation of transformer encoders is that the complexity of self-attention grows quadratically with respect to the input sequence, making it hard to train on relatively long audio samples, such as those appearing in Clotho dataset, i.e. up to 30 sec.
To address those issues, we propose a transformer AAC model that utilizes Patchout faSt Spectrogram Transformer (PaSST) \cite{passt} as the encoder. The main differences between PaSST encoder to the one proposed by Mei et al. \cite{act} is the current patch extraction method that involves: (1) A convolutional layer that extracts a feature map from the input spectrogram. (2) Decoupled time and frequency positional encoding. (3) \textit{Patchout}, where parts of the transformer’s input sequence are dropped during  training, encouraging the model to perform with an incomplete sequence.

To solve the word selection indeterminacy problem of AAC, many approaches integrate keyword information either by pre-training the encoder to predict keywords extracted from the ground truth captions \cite{keymad} or by conditioning the caption generation on input text \cite{Bart,koizumi2020audio}.
In this work, we infer AudioSet class labels from a pre-trained PaSST model and use them as guiding input text. In order to obtain text that is semantically similar to the ground truth captions and functions as a guiding text we fine-tune PaSST on audio-label pairs extracted from Clotho.

Additionally, we propose a method for using Mixup augmentation \cite{mixup} for AAC.

The remainder of this report is organized as follows: In Section 2 we discuss our proposed system. In Section 3 we present in detail our implementation. In Section 4 we present our results and in Section 5 we conclude this report.

\begin{figure*}[ht!]
  \centering
  \captionsetup{justification=centering}

  \centerline{\includegraphics[height=7cm]{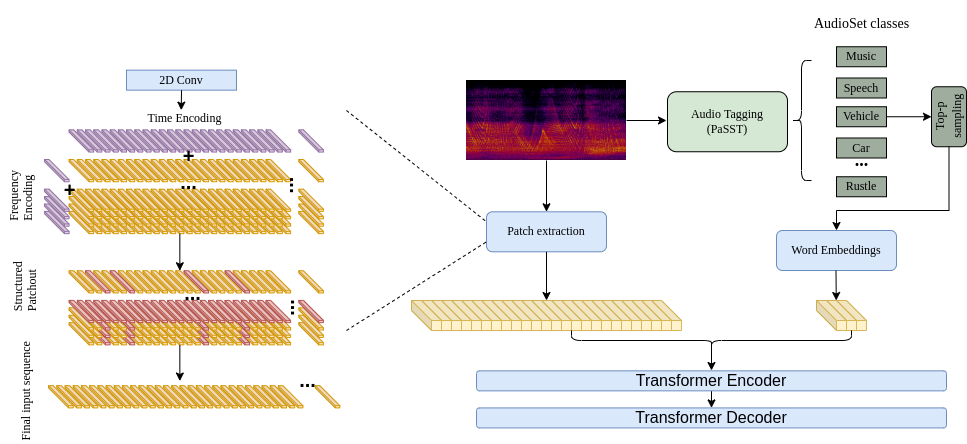}}
  \caption{\textit{The overview of our system. Blue and Green colors  indicate learned and frozen modules respectively. \\ The input to the decoder is omitted for clarity.}}
  \label{fig:results}
\end{figure*}

\section{System Description}
\label{sec:format}

The backbone architecture  of our proposed model is based on a traditional sequence-to-sequence
structure, which consists of a PaSST encoder and a Transformer decoder. 
The encoder extracts an abstract embedding sequence from the input, then this sequence is fed to the decoder, which outputs a textual description. We propose a multimodal conditioning scheme where the input sequence fed to the encoder consists of both audio and text information. The input text is obtained by a frozen, fine-tuned PaSST model.

\subsection{Patch Extraction}
Each spectrogram $\mathcal{C}$ is transformed to a sequence of input features $ X_{in} \in \mathbb{R}^{L \times h}$ through the following pipeline: (1) The spectrogram is passed through a 2-dimensional convolutional layer with kernel size 16, stride 10 and output dimension d producing feature map $X_m \in \mathbb{R}^{d \times F_m \times T_m}$. (2) Positional embeddings are added. Following the implementation of PaSST we add a frequency embedding $e_{p}^F \in \mathbb{R}^{d \times F_m \times 1}$ and a time embedding $e_{p}^T \in \mathbb{R}^{d \times 1 \times T_m}$ to inject positional information in the feature map.  (3) \textit{Strctured Patchout} \cite{passt} is applied i.e. a number of frequency bins $p_f$ and time frames $p_t$ are randomly deleted from the extracted featuremap $X_m$ resulting in $X_p \in \mathbb{R}^{d \times F_p \times T_p}$ where $F_p = F_m - p_f$ and $T_p = T_m - t_f$. (4) Finally $X_p$ is flattened to $X_{in} \in \mathbb{R}^{L \times d}$ where $L = F_p \cdot T_p $. 


\subsection{Guiding Text}

PaSST has recently achieved state-of-the-art performance in audio classification tasks \cite{passt}. Using a pre-trained PaSST model, we infer AudioSet class labels from the input audio. Each word in the label is embedded in the input space using
trainable embeddings and concatenated with the extracted patches. Similarly to \cite{Bart}, in order to make our system more robust to PaSST's prediction errors, we sample each label from the output distribution. In order to avoid the "unreliable tail" of the distribution we use Nucleus Sampling (top-p) \cite{holtzman2019curious}. During inference, we select the most probable output label instead of sampling.

Since we add word-level information to our model, we want the input label to be semantically similar to the ground truth caption, functioning as a guiding text. We observe that PaSST tends to output labels that capture the general, high-level information in the audio and not labels that are more infrequent and specific. Such labels are more likely to be semantically similar or even be present verbatim in the ground truth captions. For example, an audio clip with the caption  "\textit{A short distance away, a group of people engage in indistinguishable chatter.}" is classified as \textit{Speech} when in fact the AudioSet class \textit{Chatter} would have higher semantic accuracy. 
Based on this observation we fine-tune PaSST to predict labels that have the highest sentence similarity with the corresponding ground truth. To achieve this we use a pre-trained BERT model \footnote{https://www.sbert.net/} to encode all AudioSet classes and all captions in Clotho dataset. Then we choose the most similar class label for each caption using the cosine similarity of their BERT embedding. 
We fine-tune PaSST on these audio-label pairs, created from Clotho, by optimizing the standard binary cross entropy loss.

\begin{table*}[h]
  \centering
  \begin{tabular}{c|ccccccccc}
  \hline
  \textbf{Model} &
    \multicolumn{1}{l}{\textbf{BLUE-1}} &
    \multicolumn{1}{l}{\textbf{BLEU-2}} &
    \multicolumn{1}{l}{\textbf{BLEU-3}} &
    \multicolumn{1}{l}{\textbf{BLEU-4}} &
    \multicolumn{1}{l}{\textbf{METEOR}} &
    \multicolumn{1}{l}{\textbf{ROUGE\_L}} &
    \multicolumn{1}{l}{\textbf{CIDEr}} &
    \multicolumn{1}{l}{\textbf{SPICE}} &
    \multicolumn{1}{l}{\textbf{SPIDEr}} \\ \hline
  Baseline         & 0.555 & 0.358 & 0.239 & 0.156 & 0.164 & 0.364 & 0.358 & 0.109 & 0.233 \\ \hline
  PACT\_no\_s         & 0.576 & 0.384 & 0.261 & 0.176 & 0.166 & 0.385 & 0.453 & 0.130 & 0.292 \\
  PACT\_s             & 0.578 & 0.384 & 0.262 & 0.176 & 0.177 & 0.387 & 0.454 & 0.133 & 0.293 \\
  PACT\_s\_clr         & 0.575 & 0.384 & 0.262 & 0.174 & 0.178 & 0.386 & 0.457 & 0.133 & 0.295 \\
  PACT\_2s            & 0.579 & 0.386 & 0.262 & 0.173 & 0.178 & 0.387 & 0.457 & 0.134 & 0.296
  
  \end{tabular}
  \caption{Performances of different models on Clotho evaluation split.}
  \label{tab:my-table}
  \end{table*}

\subsection{Encoder}
In order to make use of pre-trained models, the encoder architecture
is the same as PaSST containing 12 encoder blocks and 12
heads with a hidden dimension $d=768$. 
Each layer contains two sub-layers, a multi-head self-attention layer, and a position-wise fully-connected feed-forward layer. The feed-forward network contains two linear layers with GELU 
activation function between them. Since \textit{Patcout} is used to tackle over-fitting we don't use dropout in the transformer encoder.
As shown in Figure 1 the input sequence is a concatenation of the extracted patches and the embedded guiding text.
As in ViT \cite{Vit} and AST \cite{ast}, a global learnable class token
$x_{cls} \in \mathbb{R}^{1\times d}$
is appended to the beginning of the input sequence. The classification layer of PaSST is omitted and the last hidden layer is passed
to the cross-attention layer of the decoder.

\subsection{Decoder}
The decoder consists of a word embedding layer, a
Transformer decoder block, and a linear layer. Each word from the input sequence is embedded through the word embedding layer into a vector $x_i \in \mathbb{R}^{512}$ and fed to the transformer decoder. The word vectors are extracted by a pre-trained Word2Vec model \cite{word2vec} trained on the corpus of Clotho dataset.

The transformer decoder contains 6 blocks and 8
heads with an embedding dimension of 512. Each block consists of a self-attention layer,  a cross multi-head attention layer, and a feed-forward layer. The dimension of the forward layer is 2048. The output of the encoder is adjusted through a feed-forward layer and a non-linearity and then fed to the cross multi-head attention layer of the decoder.
Positional embeddings are added and masking is applied to the input sequence.
Finally, the output of the decoder is passed to a linear layer
and a softmax function to get the output probabilities of the caption
words.

The training objective of the model is to minimize the cross entropy (CE) loss:
\begin{align}
    \mathcal{L_{CE}}(\theta) = - \frac{1}{T} \sum_{t-1}^T log p(y_t|y_{<t}, \theta)
\end{align}


where $\theta$ are the models parameters and and $y_t$ is the ground-truth word at time step $t$.

\subsection{Transfer Learning}
Since the use of external data is allowed in this task we experiment with two transfer learning schemes: utilizing a PaSST encoder trained on ImageNet and AudioSet and pre-training our model on a larger in-domain dataset. 

\subsubsection{Pre-trained encoder}
PaSST has an ImageNet pre-trained ViT as its backbone. The model is trained on AudioSet dataset and shows a powerful ability in extracting audio features
in different downstream audio pattern recognition tasks \cite{passt}. Unlike AST, PaSST uses separate embeddings for time and frequency positional encoding. 
The advantage of decoupling time and frequency embeddings
is allowing variable length inputs without fine-tuning or interpolation.

\subsubsection{Pre-training with in domain dataset}
We additionally create an enlarged in-domain dataset adding 
roughly 46,000 single caption audio samples from the AudioCaps
dataset \cite{audiocaps} and 3930 multiple caption audio samples from MACS \cite{macs} to Clotho development-validation split.
We first pre-train our model on this dataset and then fine-tune it on Clotho.

\subsection{Data Augmentation}
In order to avoid over-fitting and regularize the data we use SpecAugment \cite{specaugment}, Label Smoothing \cite{label_smoothing}, and Mixup \cite{mixup}. Since AAC is not a classification task, applying Mixup on the labels is not trivial. Unlike \cite{aac_mixup} that apply Mixup by concatenating the captions, we mix captions in the embedding space. For two sampled spectrograms $x_i$, $x_j$ and their corresponding embedded captions $y_i^e$, $y_j^e$ we apply Mixup as follows:
\begin{align*}
    x &= \lambda x_i + (1-\lambda) x_j \\
    y^e &= \lambda y_i^e + (1-\lambda) y_j^e
\end{align*}
where, $\lambda = Beta(a,a)$ and $a = 0.3$. We also mix the embedded guiding text. Despite using unmixed captions when calculating the loss, this implementation of Mixup improves the results of our model.

\section{EXPERIMENT}
\label{exp}

\subsection{Dataset}
\subsubsection{Clotho}
The latest Clotho v2 dataset contains 3839 audio clips in
the training set and 1045 audio clips in the validation
and evaluation set, respectively. The audio clips were collected from Freesound and ranged from 15 to 30 seconds. Annotators were employed through Amazon Mechanical Turk for crowdsourcing the captions. Each audio clip has five corresponding captions ranging from 8 to 20 words long. To prevent biased annotation, only the audio signal was available to the annotators. Clotho splits were created through a stratification process.

\subsubsection{AudioCaps}
AudioCaps \cite{audiocaps} is the largest audio captioning dataset containing around 46000 samples. All audio clips were sourced from AudioSet and are 10 seconds long. The annotation was conducted by crowdworkers through Amazon Mechanical Turk. Audio samples in the training set have one corresponding caption whereas validation and test audio samples have five.

\subsubsection{MACS}
MACS \cite{macs} consists of audio clips from TAU Urban Acoustic Scenes 2019 dataset. It contains 3930, 10-second long audio clips without providing subsplits. The number of captions varies from two to five captions per audio sample.

\subsection{Data pre-processing}
All audio clips are converted to 32k Hz. 
 Log mel-spectrograms are extracted using
a 1024-points Hanning window with 50\% overlap and 128 mel bins
are used as the input features. Captions are tokenized and transformed to lowercase with punctuation removed. <SOS> and <EOS> tokens are added at the beginning and the end of each caption. During pre-training on the augmented dataset, words that are not present in Clotho vocabulary are replaced with <UNK>.

\subsection{Experiment Setup}

We split the pretraining stage into two parts as in \cite{sota}. First, we freeze PaSST encoder and train on the enlarged dataset with  learning rate $1 \times 10^{-4}$.  Then the encoder is unfrozen and and a learning rate of  $1 \times 10^{-5}$ is used. During fine-tuning an initial learning  rate of $1 \times 10^{-5}$ is gradually increased to $1 \times 10^{-4}$ using linear warmup. Batch size is 32 throughout all stages. In the inference stage, we adopt beam search with a
beam width of 3.

During pre-training  we apply \textit{Patchout}, dropping 4 frequency bins and 80 time frames, and during fine-tuning on Clotho, 120 time frames. This means that during fine-tuning almost half of the input sequence is dropped resulting in a great complexity reduction. To further mitigate over-fitting we use dropout in the decoder with a rate of 0.2. The label smoothing factor is set to 0.1. 

We fine-tune a pre-trained PaSST model on the audio-label pairs we created for 1 epoch with a learning rate of $1 \times 10^{-5}$. The input guiding texts for the development-validation splits are obtained prior to training.

\section{RESULTS}

Our
submission contains the following four models:
\begin{itemize}
    \item PACT\_no\_s: This model is trained with time Patchout $p_t=80$. The input text is selected as the maximum value over PaSST's logits.
    \item PACT\_s: This model is trained with time Patchout $p_t=120$. The input text is selected using top-p sampling. 
    \item PACT\_s\_clr: Same as PACT\_s, fine-tuned on Clotho with a constant learning rate of $1\times10^{-5}$, instead of linear warm-up.
    \item PACT\_2s: This model is trained with time Patchout $p_t=120$. The input text is a concatenation of 2 AudioSet class labels selected using top-p sampling.
\end{itemize}

The performances of our submitted
models are compared with the Baseline in Table 1.

\section{CONCLUSION}
In this report, we describe our system submitted to DCASE2021
challenge, Task 6. We propose a transformer architecture with combined audio and textual conditioning.  We show that \textit{Patchout} can be effectively applied for Audio Captioning, reducing the computational complexity and optimizing performance. To obtain input guiding text that is semantically similar to the ground truth captions we fine-tune a pre-trained classification model. To solve the data scarcity problem we pre-train our model on an larger in-domain dataset and initialize the weights of our encoder with a pre-trained PaSST. The SPIDEr score of
our best model on the development-testing dataset is 0.296.
\label{sec:typestyle}


\bibliographystyle{IEEEtran}
\bibliography{template}
%
%
%
%
%
%
%
%
%

\end{sloppy}
\end{document}